\documentclass{article}

\usepackage[english]{babel}
\usepackage{cite}
\usepackage[letterpaper,top=2cm,bottom=2cm,left=3cm,right=3cm,marginparwidth=1.75cm]{geometry}

\usepackage{amsmath}
\usepackage{graphicx}
\usepackage[colorlinks=true, allcolors=blue]{hyperref}

\title{\textbf{ Digital Forensics in the Age of Smart Environments: A Survey of Recent Advancements and Challenges } }
\author{\textbf{Ahmed MohanRaj Alenezi}  \\
\date{}
\\
Independent Forensics Researcher, Auckland, New Zealand \\
Email:amam88820@gmail.com
}

\begin{document}
\maketitle


\begin{abstract}
Digital forensics in smart environments is an emerging field that deals with the investigation and analysis of digital evidence in smart devices and environments. As smart environments continue to evolve, digital forensic investigators face new challenges in retrieving, preserving, and analyzing digital evidence. At the same time, recent advancements in digital forensic tools and techniques offer promising solutions to overcome these challenges. In this survey, we examine recent advancements and challenges in digital forensics within smart environments. Specifically, we review the current state-of-the-art techniques and tools for digital forensics in smart environments and discuss their strengths and limitations. We also identify the major challenges that digital forensic investigators face in smart environments and propose potential solutions to overcome these challenges. Our survey provides a comprehensive overview of recent advancements and challenges in digital forensics in the age of smart environments, and aims to inform future research in this area.
\end{abstract}
\section{Introduction}

In recent years, smart environments have become increasingly ubiquitous in our daily lives\cite{mattern2003smart}. From smart homes and buildings to smart cities and transportation systems, the integration of smart devices and technology has led to numerous benefits, including increased efficiency, convenience, and connectivity. However, with the rise of smart environments comes new challenges for digital forensic investigators who need to retrieve, preserve, and analyze digital evidence in these environments \cite{klein2008smart, martins2012smart}\cite{mccreary2016contextual}.

Digital forensics is the process of identifying, preserving, analyzing, and presenting digital evidence in a manner that is admissible in a court of law \cite{teing2017forensic,taylor2011forensic}. Digital forensics plays a crucial role in investigating cybercrimes\cite{sun2015survey,atlam2020security}, and has become increasingly important as more and more devices become connected to the internet moreso in the Internet of things. In smart environments, digital forensic investigators face unique challenges that require specialized techniques and tools to overcome and there has been approaches on conducting security and digital forensics in connected IoT environments \cite{kebande2016generic}.

One of the key challenges of digital forensics in smart environments is the sheer number and variety of devices that are connected to the internet. Smart environments often consist of a multitude of devices, including sensors, cameras, smart speakers, and other internet of things (IoT) devices \cite{plachkinova2016emerging,martins2012smart,hou2019survey,sharma2020internet,kebande2016towards,karie2016generic}. Each of these devices can generate a large amount of digital evidence that needs to be collected and analyzed. Furthermore, these devices often use different operating systems, communication protocols, and storage formats, which can complicate the process of digital forensic analysis.

Despite these challenges, recent advancements in digital forensic techniques and tools offer promising solutions for investigating digital evidence in smart environments. For example, new tools have been developed to collect data from IoT devices, such as smart home devices and wearables. These tools can help investigators retrieve digital evidence from a variety of sources and formats, and can provide insights into the activity and behavior of users in smart environments.

\section{Background}

Digital forensics and smart environments are two rapidly growing fields in the world of technology. Digital forensics refers to the process of identifying, preserving, analyzing, and presenting digital evidence in a court of law, while smart environments are physical spaces that are equipped with intelligent technology and connected devices. With the rise of smart environments, digital forensic investigators are faced with new challenges in retrieving, preserving, and analyzing digital evidence. This requires specialized techniques and tools that can handle the diverse range of devices and data sources found in smart environments. In this background study, we will explore the intersection of these two fields and discuss the key challenges and advancements in digital forensics in smart environments.

\subsection{Digital Forensics}

Digital forensics  involves a variety of techniques and methodologies to acquire, preserve, and analyze digital evidence \cite{kebande2020mapping,guo2012forensic,taylor2011forensic}. In the acquisition phase, investigators use specialized tools and software to create a forensic image of a device or storage medium. This image is an exact copy of the original data and is used for analysis and preservation purposes. Preservation is a critical phase of digital forensics, as any alteration or modification to the data can render it inadmissible in court \cite{damshenas2012forensics,alex2017forensics, holobinko2012forensic,kigwana2017proposed}.

\begin{figure}
    \centering
    \includegraphics[width=0.5\textwidth]{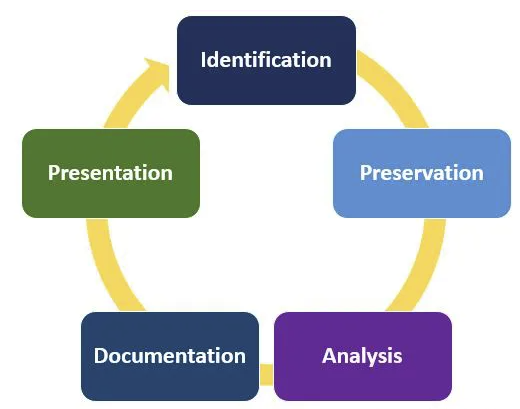}
    \caption{Digital forensic process}
    \label{fig:example1}
\end{figure}

Once the data has been acquired and preserved, the analysis phase begins. In this phase, digital forensic investigators use a range of tools and techniques to extract information from the data and identify relevant evidence \cite{sree2020data,meffert2017forensic}. This can involve keyword searches, data carving, metadata analysis, and other methods \cite{muhammad2017smart,adedayo2016big}. An example of forensic process is shown in \ref{fig:example1}.

Digital forensics is used in a variety of settings, including criminal investigations, civil litigation, and corporate investigations \cite{teing2017forensic,holobinko2012forensic, damshenas2012forensics}. In criminal investigations, digital forensics can be used to identify suspects, track their online activities, and retrieve evidence from digital devices \cite{kebande2022finite, kebande2022industrial}. In civil litigation, digital forensics can be used to uncover evidence of intellectual property theft, breach of contract, and other legal violations \cite{zhang2020digital,zawoad2015ocf,murff2011digital,krishnan2021interplay,danielsson2004need}. In corporate investigations, digital forensics can be used to investigate data breaches, employee misconduct, and other security incidents \cite{kebande2014cloud,kebande2015functional}.

In recent years, the use of digital forensics has become increasingly important in the context of cybercrime. With the rise of cyberattacks, digital forensics plays a critical role in identifying the source of attacks, tracking the activities of cybercriminals, and recovering stolen data\cite{kebande2017iot,zawoad2015digital,ieong2006forza, agarwal2011systematic, baror2021framework,baror2022conceptual}.

Overall, digital forensics is a complex and constantly evolving field that requires specialized knowledge and skills. As technology continues to advance, digital forensic investigators must stay up to date with the latest tools and techniques in order to effectively retrieve, preserve, and analyze digital evidence \cite{alex2017forensics,thethi2014digital, karie2019diverging, alqahtany2015cloud, almulla2013cloud,kigwana2017towards}.

\subsection{Smart Environments}

Smart environments refer to physical spaces that are equipped with intelligent technology and connected devices, such as sensors, cameras, and other IoT devices \cite{holmquist2004building,le2001smart,kebande2015towards}. These environments are designed to be more efficient, comfortable, and secure for their occupants, while also providing new opportunities for businesses and organizations\cite{youngblood2005automation, augusto2018user,kebande2018digital}.

\begin{figure}
    \centering
    \includegraphics[width=0.5\textwidth]{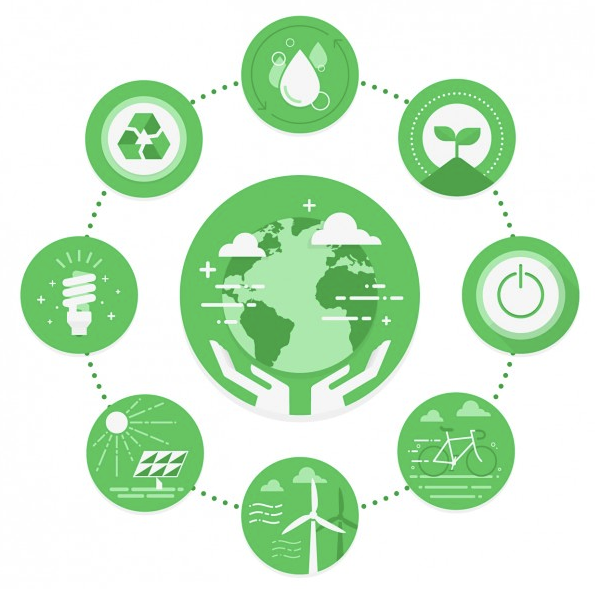}
    \caption{Smart environment example}
    \label{fig:example}
\end{figure}

Smart environments are becoming increasingly popular in a variety of settings, including homes, offices, schools, hospitals, and public spaces \cite{augusto2010ambient,poux2019smart,o2012smart}.. In a smart home, for example, devices can be connected and controlled remotely through a smartphone app, allowing users to control lighting, heating, and security systems from anywhere \cite{karie2021review}.. In a smart office, sensors can monitor temperature, humidity, and occupancy levels to optimize energy usage and improve productivity\cite{cook2004smart,holthaus2016address,fortino2014middlewares,karie2018knowledge}. Example of Smart environment is shown in \ref{fig:example}.

One of the key benefits of smart environments is their ability to generate and collect large amounts of data\cite{cook2007smart,dey2000context,lamnatou2022smart}. This data can be used to improve the performance and efficiency of the environment, as well as to gain insights into user behavior and preferences. However, this also presents new challenges for data security and privacy\cite{benevolo2016smart,guan2012smart,das2005designing}.

In smart environments, digital forensics plays a critical role in retrieving, preserving, and analyzing digital evidence. With the large number of devices and data sources in a smart environment, investigators must have specialized knowledge and tools to effectively retrieve and analyze evidence \cite{babun2018iotdots,ross2020security,park2018research,kebande2018adding}.. Additionally, the distributed nature of smart environments, with data stored in remote servers and virtual environments, presents new challenges for digital forensics. Smart environments are a rapidly growing field that presents new opportunities and challenges for businesses, organizations, and individuals. As these environments become more prevalent, it is essential to have effective digital forensics techniques and tools in place to ensure the security and privacy of the data generated and collected \cite{sree2020data}.

\section{Survey of Recent Advancements} 

As smart environments continue to grow in popularity and complexity, digital forensics techniques are evolving to meet the unique challenges posed by these environments. In recent years, there have been significant advancements in the field of digital forensics, particularly in the area of smart environments.

\subsection{Tool development} 
One major advancement has been the development of new tools and techniques for analyzing data from a wide range of devices and sources. With the proliferation of IoT devices in smart environments, digital forensics investigators must be able to retrieve and analyze data from sensors, cameras, and other sources. To meet this challenge, new tools and techniques have been developed, including network forensics, cloud forensics, and memory forensics\cite{lopez2018smart,desolda2017empowering,robles2010context}. As smart environments become increasingly complex and diverse, digital forensics investigators must be able to retrieve and analyze data from a wide range of devices and sources, including IoT sensors, mobile devices, cloud-based services, and more. To meet this challenge, new tools and techniques have been developed to help investigators analyze and interpret the large volumes of data that are generated in these environments.

One important development has been the emergence of network forensics tools and techniques. Network forensics involves the analysis of network traffic to identify and analyze digital evidence. In smart environments, network forensics can be used to identify unauthorized access to devices or data, as well as to track the movement of data through the environment. Network forensics tools and techniques can help investigators identify and analyze network traffic patterns, extract data from network packets, and reconstruct network sessions.

Cloud forensics has also emerged as a critical area of digital forensics in the age of smart environments. Cloud forensics involves the analysis of data stored in cloud-based services and virtual environments \cite{ruan2011cloud,ruan2013cloud,zawoad2013cloud,simou2014cloud,simou2016survey,dykstra2011understanding,kebande2018novel}. With the increasing use of cloud-based services in smart environments, cloud forensics has become an essential component of digital forensics investigations. Cloud forensics tools and techniques can help investigators extract data from cloud storage services, analyze virtual machine images, and reconstruct user activity in cloud-based environments.

\begin{table}[ht]
\centering
\caption{Development of New Tools and Techniques for Analyzing Data in Smart Environments}
\label{tab:dev}
\begin{tabular}{|l|p{10cm}|}
\hline
\textbf{Development} & \textbf{Description} \\ \hline
Network forensics & Involves the analysis of network traffic to identify and analyze digital evidence. Can be used to identify unauthorized access to devices or data, as well as to track the movement of data through the environment. Tools and techniques can help identify and analyze network traffic patterns, extract data from network packets, and reconstruct network sessions. \\ \hline
Cloud forensics & Involves the analysis of data stored in cloud-based services and virtual environments. Can be used to extract data from cloud storage services, analyze virtual machine images, and reconstruct user activity in cloud-based environments. \\ \hline
Memory forensics & Involves the analysis of data stored in a computer's memory. Can be used to uncover evidence of malware, network intrusions, and other security incidents. Can be used to analyze the memory of IoT devices, mobile devices, and other systems to identify evidence of security incidents or cybercrime. \\ \hline
\end{tabular}
\end{table}

\subsection{Network Forensics }
Network forensics involves the analysis of network traffic to identify and analyze digital evidence. In smart environments, network forensics can be used to identify unauthorized access to devices or data, as well as to track the movement of data through the environment \cite{khan2016network}. Cloud forensics involves the analysis of data stored in cloud-based services and virtual environments. With the increasing use of cloud-based services in smart environments, cloud forensics has become an essential component of digital forensics investigations. Memory forensics involves the analysis of data stored in a computer's memory, and can be used to uncover evidence of malware, network intrusions, and other security incidents \cite{corey2002network, meghanathan2010tools}. .

Network forensics is a branch of digital forensics that focuses on the analysis of network traffic to identify and analyze digital evidence. Network forensics is particularly useful in the investigation of cybercrimes and security incidents in smart environments, as it allows investigators to identify unauthorized access to devices or data, track the movement of data through the environment, and identify sources of network-based attacks\cite{hunt2012network, messier2017network,karie2016building,karie2019importance }.

Network forensics can be used to identify a range of different types of attacks, including denial-of-service (DoS) attacks, distributed denial-of-service (DDoS) attacks, and network-based malware infections. By analyzing network traffic patterns, investigators can identify unusual or suspicious traffic flows, such as a large volume of traffic originating from a single source or unusual traffic patterns that deviate from normal behavior.

Network forensics tools and techniques can help extract data from network packets, reconstruct network sessions, and analyze network traffic patterns to identify evidence of cybercrime. Network packet capture tools such as Wireshark, tcpdump, and Snort can be used to capture and analyze network traffic, while network flow analysis tools like NetFlow and sFlow can help identify the source, destination, and type of traffic on a network as is shown in Table \ref{tab:net}.

\begin{table}[ht]
\centering
\caption{Network Forensics Tools and Techniques}
\label{tab:net}
\begin{tabular}{|l|p{10cm}|}
\hline
\textbf{Tool/Technique} & \textbf{Description} \\ \hline
Packet capture tools & Tools such as Wireshark, tcpdump, and Snort can be used to capture and analyze network traffic. They allow investigators to identify unusual or suspicious traffic flows, such as a large volume of traffic originating from a single source or unusual traffic patterns that deviate from normal behavior. \\ \hline
Network flow analysis & Tools like NetFlow and sFlow can help identify the source, destination, and type of traffic on a network. They can be used to detect DoS and DDoS attacks, malware infections, and other types of network-based attacks. \\ \hline
Session reconstruction & Tools such as NetworkMiner and NetWitness can be used to reconstruct network sessions, allowing investigators to analyze the data transferred during a session and identify any anomalies or evidence of cybercrime. \\ \hline
\end{tabular}
\end{table}

Network forensics is an essential tool for investigating security incidents and cybercrime in smart environments. With the growing number of devices connected to networks in smart environments, network forensics is becoming increasingly important for identifying and responding to security incidents and protecting sensitive data.
\subsection{Artificial Intelligence } 
Another major advancement in digital forensics has been the use of artificial intelligence (AI) and machine learning (ML) techniques to analyze data and identify patterns. In smart environments, AI and ML can be used to analyze large amounts of data from sensors and other sources, and to identify anomalous behavior or security incidents. With the increasing volume of data generated in smart environments, traditional manual methods of digital forensics can be time-consuming and challenging. AI and ML techniques can be used to automate parts of the digital forensic process and help investigators identify patterns and anomalies in large datasets.

AI and ML techniques can be used for a range of different purposes in digital forensics, including image and audio analysis, text analysis, and behavioral analysis. For example, AI and ML algorithms can be trained to recognize faces or voices in images or audio recordings, helping investigators identify suspects or potential sources of evidence.

In text analysis, AI and ML can be used to identify keywords, patterns, and sentiment in large volumes of text data, such as chat logs or email messages. This can help investigators identify potential leads or evidence related to a case.

In behavioral analysis, AI and ML can be used to identify unusual or suspicious patterns of behavior in large datasets, such as user activity logs or network traffic. This can help investigators identify potential threats or incidents of cybercrime.

However, there are also challenges associated with the use of AI and ML in digital forensics. One challenge is the need for large volumes of training data to train the algorithms effectively. Another challenge is the potential for bias in the algorithms, which can lead to errors or inaccurate results. It is important for digital forensics investigators to be aware of these challenges and to use AI and ML techniques appropriately. 

The use of AI and ML techniques in digital forensics is a major advancement that has the potential to improve the speed and accuracy of digital forensic investigations. However, there are also challenges associated with the use of these techniques, and it is important for investigators to use them appropriately and be aware of their limitations as is shown in \ref{tab:AI_ML}.

\begin{table}[htbp]
    \caption{Techniques in Digital Forensics}
    \label{tab:AI_ML}
    \centering
    \begin{tabular}{|p{3cm}|p{6cm}|p{4cm}|}
    \hline
    \textbf{Technique} & \textbf{Application} & \textbf{Advantages} \\
    \hline
    Image and audio analysis & Identification of faces, voices, and objects in images or audio recordings & Helps identify suspects or potential sources of evidence \\
    \hline
    Text analysis & Identification of keywords, patterns, and sentiment in large volumes of text data such as chat logs or email messages & Helps investigators identify potential leads or evidence related to a case \\
    \hline
    Behavioral analysis & Identification of unusual or suspicious patterns of behavior in large datasets such as user activity logs or network traffic & Helps identify potential threats or incidents of cybercrime \\
    \hline
    \end{tabular}
\end{table}

\subsection{Forensic Readiness } 
Finally, there have been advancements in the area of forensic readiness planning, which involves preparing an organization or environment for the possibility of a digital forensics investigation. Forensic readiness planning can include measures such as data encryption, user access controls, and the implementation of logging and auditing procedures.

Forensic readiness is a proactive approach taken by organizations to ensure that they are prepared to collect, preserve, and analyze digital evidence in the event of a security incident or cybercrime. In other words, forensic readiness is about being prepared to conduct a digital forensic investigation when it is needed \cite{danielsson2004need,kebande2019cfraas}.
.

Digital forensic investigations are a critical component of incident response. By being forensic-ready, an organization can significantly reduce the time and cost associated with conducting a digital forensic investigation. This is because forensic readiness involves putting in place policies, procedures, and tools that enable the rapid collection and analysis of digital evidence.\cite{mouhtaropoulos2014digital,kebande2015towards}.

Forensic readiness involves a wide range of activities, including the identification of critical data sources and the development of procedures for collecting and preserving digital evidence. Organizations that are forensic-ready will also have trained personnel who are able to respond quickly and effectively to a security incident\cite{alenezi2019experts,kebande2016requirements}.
.

One of the key components of forensic readiness is the establishment of a digital forensic laboratory. A digital forensic laboratory is a secure facility where digital evidence can be collected, analyzed, and stored. A digital forensic laboratory should be equipped with the necessary hardware, software, and personnel to conduct digital forensic investigations\cite{park2018research,kebande2017novel,kebandeadding}.

Another important aspect of forensic readiness is the development of forensic tools and techniques. Forensic tools and techniques are used to collect and analyze digital evidence. These tools can range from simple software tools that enable the collection of digital evidence from a single device to complex systems that can collect and analyze data from multiple sources simultaneously.

Forensic readiness also involves the development of policies and procedures for collecting and preserving digital evidence. These policies and procedures should be designed to ensure that digital evidence is collected and preserved in a forensically sound manner. This means that the evidence must be collected and preserved in a way that maintains its integrity, so that it can be used as evidence in court if necessary\cite{mouhtaropoulos2014digital,kebande2015towards,kebande2015functional,kebande2016towards,kebande2019cfraas}.

In addition to developing policies and procedures for digital evidence collection and preservation, organizations that are forensic-ready should also have policies and procedures in place for incident response. Incident response procedures should be designed to ensure that security incidents are detected and responded to quickly and effectively. This includes procedures for identifying the type of incident, containing the incident, and conducting a digital forensic investigation.

Forensic readiness is a proactive approach that organizations can take to ensure that they are prepared to conduct digital forensic investigations in the event of a security incident or cybercrime. Forensic readiness involves the development of policies, procedures, and tools for collecting, preserving, and analyzing digital evidence. By being forensic-ready, organizations can significantly reduce the time and cost associated with conducting digital forensic investigations, and improve their overall cybersecurity posture.

There have been significant advancements in digital forensics in the age of smart environments, including new tools and techniques for analyzing data, the use of AI and ML, and the implementation of forensic readiness planning measures. These advancements are essential for ensuring the security and privacy of data in smart environments, and for investigating security incidents and cybercrime as is shown in \ref{table:forensic-readiness}.

\begin{table}[htbp]
    \caption{Digital forensics Readiness challenges}
    \label{tab:AI_ML}
    \centering
\begin{tabular}{|p{2.5cm}|p{10cm}|}
\hline
\textbf{Aspect} & \textbf{Description} \\
Forensic readiness & The process of preparing an organization to be able to effectively respond to digital forensic incidents. This involves developing policies, procedures, and technical capabilities to collect and preserve digital evidence in a forensically sound manner\cite{tan2001forensic,rowlingson2004ten,kebande2019cfraas}.\\
\hline
Forensic readiness frameworks & Frameworks such as the UK Association of Chief Police Officers (ACPO) guidelines and the National Institute of Standards and Technology (NIST) guidelines provide guidance on best practices for forensic readiness. These frameworks cover areas such as policy development, staff training, evidence collection, and incident response \cite{owen2011analysis}. \\
\hline
Benefits of forensic readiness & Forensic readiness can help organizations respond more effectively to digital forensic incidents, reduce the risk of evidence being lost or compromised, and improve the quality of evidence collected. It can also help organizations comply with legal and regulatory requirements related to digital forensics \cite{kebande2019comparative,kebande2022finite}. \\ 
\hline
Challenges of forensic readiness & Implementing forensic readiness can be challenging and time-consuming, particularly for smaller organizations with limited resources. Maintaining and updating forensic readiness policies and procedures can also be a challenge, as technology and threats continue to evolve\cite{kebande2015adding,kebande2018novel,kebande2017novel, sim2022argus,simou2014cloud}.\\
\hline
Research on forensic readiness & Researchers such as  Kebande et al  \cite{kebande2015adding,kebande2018novel,kebande2017novel, kebande2018forensic} have conducted studies on forensic readiness, exploring topics such as the effectiveness of forensic readiness frameworks and the challenges of implementing forensic readiness in different organizational contexts. \\ 
\hline
Cloud forensic readiness & Cloud forensic readiness involves developing policies, procedures, and technical capabilities to collect and preserve digital evidence in cloud environments\cite{ruan2011cloud, ruan2013cloud}. . This includes addressing challenges such as data privacy, data ownership, and the distributed nature of cloud environments \cite{kebande2018novel,kebande2017novel, kebande2018forensic}. \\
\hline
\end{tabular}
\caption{Summary of Forensic Readiness Aspects}
\label{table:forensic-readiness}
\end{table}

\section{Challenges of Digital Forensics in the Age of Smart
Environments}

As the use of smart environments continues to grow and expand, digital forensics faces several challenges in its efforts to keep pace with the evolving landscape of technology. The increasing complexity and diversity of digital devices, systems, and networks, coupled with the sheer volume of data generated in smart environments, have created new challenges for digital forensic investigators. These challenges include technical, legal, and ethical issues that require careful consideration and innovative solutions.
\begin{itemize}
    \item 
\subsection{General Challenges} 
As smart environments continue to grow and evolve, digital forensics is facing several challenges in keeping up with the ever-increasing volume and complexity of data. Some of the major challenges facing digital forensics \cite{alenezi2023digital} in the age of smart environments are:

 \item Data volume and complexity: With the increasing volume of data generated in smart environments, the amount of digital evidence that needs to be analyzed is growing exponentially. This requires digital forensic investigators to use more advanced tools and techniques to manage and analyze large datasets.

 \item Data heterogeneity: Smart environments generate data from a wide range of devices and sources, such as sensors, IoT devices, social media, and cloud services. The data can be in different formats and structures, making it challenging for digital forensic investigators to collect, analyze, and correlate the data to reconstruct the events.

 \item Data privacy and security: Smart environments often involve personal and sensitive data, such as location data, health data, and financial data. Protecting the privacy and security of this data is crucial, but it can also pose challenges for digital forensic investigators who need access to the data to conduct their investigations \cite{nixon2004security,marky2020don}.

 \item Legal and regulatory issues: Digital forensic investigations in smart environments often involve legal and regulatory issues, such as data ownership, data retention, and privacy laws. Digital forensic investigators need to be aware of these issues and comply with the relevant laws and regulations.

 \item Lack of standards and best practices: There is a lack of standardized procedures and best practices for conducting digital forensic investigations in smart environments. This can lead to inconsistencies and errors in the investigation process and can make it difficult to share and compare results between investigators.

 \item Technical challenges: Smart environments often use advanced technologies, such as encryption, virtualization, and cloud computing, which can pose technical challenges for digital forensic investigations. These challenges include identifying and recovering deleted or corrupted data, analyzing encrypted data, and preserving the integrity of evidence in a cloud environment \cite{kebande2016generic}.
\end{itemize}

Digital forensics in the age of smart environments faces several challenges, including the volume and complexity of data, data heterogeneity, data privacy and security, legal and regulatory issues, lack of standards and best practices, and technical challenges. Addressing these challenges requires a combination of technical expertise, legal knowledge, and best practices that can adapt to the evolving landscape of smart environments.

\subsection{Digital Forensics Challenges} 

Forensic challenges in the age of smart environments are numerous and complex\cite{vincze2016challenges,karie2017taxonomy,manral2019systematic,rigworoindustrial}. As digital devices and smart environments become more ubiquitous, the volume of digital data and the complexity of the data sources also increase. This creates several challenges for digital forensic investigators who must collect, preserve, analyze, and present digital evidence in a court of law \cite{vincze2016challenges,kebande2015obfuscating, montasari2020digital,kebande2016mobile}.

One of the primary challenges in digital forensics is the lack of standardization in data formats and communication protocols used by different smart devices and platforms. This can make it difficult for investigators to collect and analyze data from different sources and devices, leading to inconsistencies and inaccuracies in the analysis of digital evidence\cite{caviglione2017future, casino2022research}.

Another challenge is the use of encryption and other security measures in smart devices and platforms. While encryption is an essential tool for protecting user privacy and securing sensitive data, it also presents challenges for digital forensic investigators who must obtain access to encrypted data to collect and analyze evidence\cite{mohay2005technical, kebande2015adding,}.

The increasing complexity and sophistication of cyberattacks also pose significant challenges for digital forensics investigators. Attackers can use a range of techniques to hide their tracks and cover their digital footprints, making it difficult for investigators to identify and attribute cybercrimes\cite{baig2017future,kebande2018uml, kebande2016towards}.

The rapid pace of technological innovation in smart environments also presents challenges for digital forensic investigators who must stay up-to-date with new technologies and tools to effectively investigate cybercrimes. Failure to keep pace with technological advancements can lead to a lack of knowledge and skills in digital forensic investigations, resulting in incomplete or inaccurate analyses of digital evidence\cite{al2013challenges}.

Additionally, the storage and retrieval of large amounts of digital data can be challenging, requiring advanced storage and processing capabilities to handle massive volumes of data generated by smart environments. This can create logistical and financial challenges for organizations that need to invest in the necessary infrastructure to handle the data\cite{lillis2016current}.

Finally, there are also ethical and legal challenges associated with digital forensics investigations in smart environments. For example, the collection and use of personal data and privacy concerns must be carefully balanced against the need to investigate and prosecute cybercrimes.

\subsection{Discussions} 

The study on Digital Forensics in the Age of Smart Environments: A Survey of Recent Advancements and Challenges provides a comprehensive overview of the recent advancements and challenges in the field of digital forensics. The study highlights the significant impact of smart environments on digital forensics, including the increasing volume and complexity of data generated in these environments and the need for new tools and techniques to manage and analyze this data.

One of the strengths of this study is the extensive literature review, which provides a comprehensive overview of the existing research on digital forensics in smart environments. The authors have included a wide range of studies, which ensures that the review is both thorough and informative.

Another strength of the study is the clear presentation of the key advancements and challenges in digital forensics. The authors have presented these in a logical and well-structured manner, making it easy for readers to understand the key issues and to identify areas where further research is needed.

However, one limitation of the study is that it is focused primarily on advancements and challenges related to traditional digital forensics, such as the analysis of digital devices and networks. While the authors briefly mention the challenges of forensic investigations in the context of the Internet of Things (IoT) and cloud computing, these areas could have been explored in more detail.

Furthermore, while the study provides a comprehensive overview of the existing research on digital forensics in smart environments, it does not present any new research findings or original data. This means that the study is limited to providing a synthesis of existing research, rather than contributing new insights or perspectives.

Consequently, the study on Digital Forensics in the Age of Smart Environments: A Survey of Recent Advancements and Challenges is a valuable contribution to the field of digital forensics. The study provides a thorough and well-organized review of the existing research on digital forensics in smart environments, highlighting the key advancements and challenges in this area. While there are some limitations to the study, such as the limited exploration of IoT and cloud computing, the study serves as an important resource for researchers and practitioners in the field of digital forensics.

One of the strengths of this study is its comprehensive survey of recent advancements and challenges in digital forensics in the age of smart environments. The authors have identified key areas of development, such as the use of AI and ML techniques and the importance of forensic readiness, and have provided an in-depth analysis of their potential impact on digital forensics investigations.

Moreover, the study has highlighted the challenges that digital forensics investigators face in smart environments. These challenges include the increasing volume of data generated by smart devices, the complexity of the data, and the potential for data tampering or destruction. By addressing these challenges, the authors have contributed to a better understanding of the unique challenges posed by smart environments and have provided insights on how to mitigate them.

Another strength of this study is its relevance to current and future trends in digital forensics and identification of threats in smart IoT environments \cite{kebande2020internet,kim2020review,ullah2019cyber}.. As smart environments continue to grow in popularity and complexity, digital forensics investigators will face increasing challenges in collecting and analyzing digital evidence. The study provides valuable insights and recommendations that can help investigators stay ahead of the curve and effectively address these challenges.

However, one limitation of this study is its focus on recent advancements and challenges in digital forensics, which may quickly become outdated as technology and smart environments continue to evolve. Additionally, the study does not provide a detailed analysis of specific smart environments, such as the Internet of Things (IoT) or smart homes, which may have unique challenges that require further investigation.

Overall, the study provides a valuable contribution to the field of digital forensics by highlighting the unique challenges posed by smart environments and identifying key areas of development. The insights and recommendations provided can help digital forensics investigators stay up-to-date with the latest advancements and effectively address the challenges of investigating in smart environments. However, further research is needed to address the specific challenges posed by different types of smart environments and to ensure that digital forensics continues to evolve and adapt to the changing technological landscape.
\section{Conclusions} 

Moving forward, there is a need for continued research and development in digital forensics in smart environments to address the challenges outlined in this survey. One area of focus could be the development of new techniques for collecting and analyzing data in real-time to keep up with the velocity of data generated by smart technologies. Additionally, the development of more robust and unbiased AI and ML algorithms is crucial to improving the accuracy and reliability of digital forensic investigations. There is also a need for continued collaboration between researchers, industry professionals, and policymakers to address legal and privacy concerns surrounding the use of data in smart environments.

Overall, as smart technologies continue to become more prevalent in our daily lives, it is essential to continue advancing digital forensics techniques to keep up with the ever-growing volume and complexity of digital data generated.

\section{Future Work} 

Moving forward, there is a need for continued research and development in digital forensics in smart environments to address the challenges outlined in this survey. One area of focus could be the development of new techniques for collecting and analyzing data in real-time to keep up with the velocity of data generated by smart technologies. Additionally, the development of more robust and unbiased AI and ML algorithms is crucial to improving the accuracy and reliability of digital forensic investigations. There is also a need for continued collaboration between researchers, industry professionals, and policymakers to address legal and privacy concerns surrounding the use of data in smart environments.

Overall, as smart technologies continue to become more prevalent in our daily lives, it is essential to continue advancing digital forensics techniques to keep up with the ever-growing volume and complexity of digital data generated.

\bibliographystyle{ieeetr}
\bibliography{sample}

\end{document}